# AB Electronic Tubes and Quasi-Superconductivity at Room Temperature*

**Alexander Bolonkin**
C&R, 1310 Avenue R, #F-6, Brooklyn, NY 11229, USA
T/F 718-339-4563, aBolonkin@juno.com, aBolonkin@gmail.com, http://Bolonkin.narod.ru

**Abstract**

   Author offers and researches a new idea – filling tubes by electronic gases. He shows: If the insulating envelope (cover) of the tube is charged positively, the electrons within the tube are not attracted to covering. Tube (as a whole) remains a neutral (uncharged) body. The electron gas in the tube has very low density and very high conductivity, close to superconductivity. If we take the density (pressure) of electron gas as equal to atmospheric pressure, the thickness of insulator film may be very small and the resulting tube is very light.
   Author shows the offered tubes can be applied to many technical fields. For example:
(1) Transfer of energy over very long distance with very small electric losses. (2) Design of cheap high altitude electric lines without masts. (3) Transfer of energy from one continent to another continent through the ionosphere. (4) Transfer of a plasma beam (which can convey thrust and energy) from Earth's surface to a space ship. (5) Observation of the sky by telescope without atmospheric hindrances. (6) Dirigibles (air balloons) of the highest lift force. (7) Increasing of gun range severalfold. (8) Transfer of matter. And so on.
-------------------
*Presented to http://arxiv.org on 8 April, 2008.

## Introduction

   The author's first innovations in electrostatic applications were developed in 1982-1983 [1]-[2]. Later the series articles of this topic were published in [3]-[17]. In particular, in the work [3] was developed theory of electronic gas and its application to building (without space flight!) inflatable electrostatic space tower up to the stationary orbit of Earth's satellite (GEO).
   In given work this theory applied to special inflatable electronic tubes made from thin insulator film. It is shown the charged tube filled by electron gas is electrically neutral, that can has a high internal pressure of the electron gas.
   The main property of AB electronic tube is a very low electric resistance because electrons have small friction on tube wall. (In conventional solid (metal) conductors, the electrons strike against the immobile ions located in the full volume of the conductor.). The abnormally low electric resistance was found along the lateral axis only in nanotubes (they have a tube structure!). In theory, metallic nanotubes can have an electric current density (along the axis) more than 1,000 times greater than metals such as silver and copper. Nanotubes have excellent heat conductivity along axis up 6000 W/m·K. Copper, by contrast, has only 385 W/m·K. The electronic tubes explain why there is this effect. Nanotubes have the tube structure and electrons can free move along axis (they have only a friction on a tube wall).
   More over, the moving electrons produce the magnetic field. The author shows - this magnetic field presses against the electron gas. When this magnetic pressure equals the electrostatic pressure, the electron gas may not remain in contact with the tube walls and their friction losses. The electron tube effectively becomes a superconductor for any surrounding temperature, even higher than room temperature! Author derives conditions for it and shows how we can significantly decrease the electric resistance.

## Description, Innovations, and Applications.

**Description**. An electronic **AB-Tube** is a tube filled by electron gas (fig.1). Electron gas is the lightest gas known in nature, far lighter than hydrogen. Therefore, tubes filled with this gas have the maximum possible lift force in atmosphere (equal essentially to the lift force of vacuum). The applications of electron gas are based on one little-known fact – the electrons located within a cylindrical tube having a positively charged cover (envelope) are in neutral-charge conditions – the total attractive force of the positive envelope plus negative contents equals zero. That means the electrons do not adhere to positive charged tube cover. They will freely fly into an **AB-Tube.** It is known, if the Earth (or other planet) would have, despite the massive pressures there, an empty space in Earth's very core, any matter in this (hypothetical!) cavity would be in a state of weightlessness (free fall). All around, attractions balance, leaving no vector 'down'.

Analogously, that means the **AB-Tube** is a conductor of electricity. Under electric tension (voltage) the electrons will collectively move without internal friction, with no vector 'down' to the walls, where friction might lie. In contrast to movement of electrons into metal (where moving electrons impact against a motionless ion grate). In the **AB-Tube** we have only electron friction about the tube wall. This friction is significantly less than the friction electrons would experience against ionic structures—and therefore so is the electrical resistance.

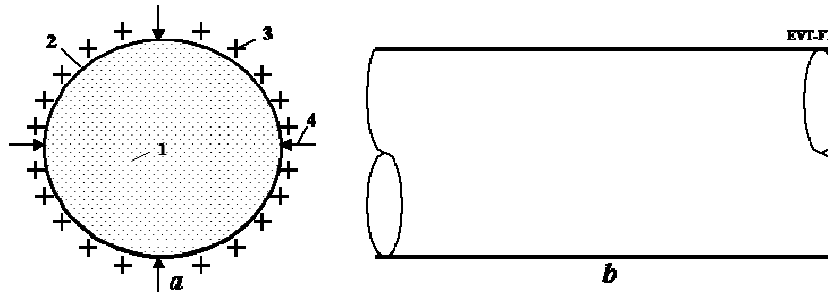

**Fig.1**. Electronic vacuum **AB-Tube**. *a*) Cross-section of tube. *b*) Side view. *Notation*: 1 – Internal part of tube filled by free electrons; 2 – insulator envelope of tube; 3 – positive charges on the outer surface of envelope (over this may be an additional film-insulator); 4 – atmospheric pressure.

When the density of electron gas equals $n = 1.65 \times 10^{16}/r$ 1/m$^3$ (where $r$ is radius of tube, m), the electron gas has pressure equals atmospheric pressure 1 atm (see research below). In this case the tube cover may be a very thin—though well-sealed-- insulator film. The outer surface of this film is charged positively by static charges equal the electron charges and **AB-Tube** is thus an electrically neutral body.

Moreover, when electrons move into the **AB-Tube**, the electric current produces a magnetic field (fig.2). This magnetic field compresses the electron cord and decreases the contact (and friction, electric resistance) electrons to tube walls. In the theoretical section is received a simple relation between the electric **current** and linear tube charge when the magnetic pressure equals to electron gas pressure $i = c\tau$ (where $i$ is electric **current**, A; $c = 3 \times 10^8$ m/s – is the light speed; $\tau$ is tube linear electric charge, C/m). In this case the electron friction equals zero and **AB-Tube** becomes **superconductive at any outer temperature**. Unfortunately, this condition requests the electron speed equals the light speed. It is, however, no problem to set the electron speed very close to light speed. That means we can make the electric conductivity of **AB-Tube**s very close to superconductivity almost regardless of the outer temperature.

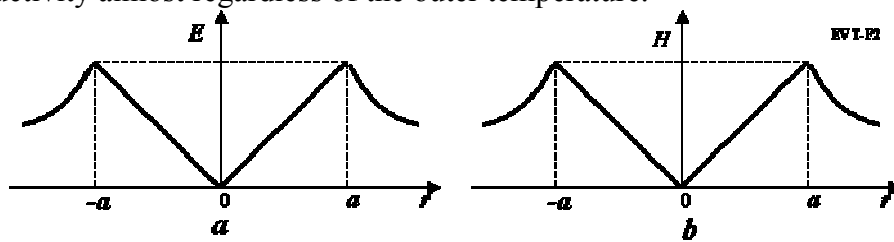

**Fig. 2.** Electrostatic and magnetic intensity into **AB-Tube**. *a*) Electrostatic intensity (pressure) via tube radius. *b*) Magnetic intensity (pressure) from electric **current** versus rube radius.

Example of electric line using the electron tubes is presented in fig.3. That consists the conventional constant-voltage electric generator 1, two nets 2 accelerating the electrons, two nets 4 braking the electrons and the electron tubes 3 connecting the electric source and customer.

A matter worthy of your special attention: The insertion of electric energy into electronic **AB-Tubes** is accomplished not by the conventional method (direct contact to contactor), but the acceleration of electron flow by electric field between two nets. Accordingly, the extraction of electric energy from electron flow is also by means of the two nets by braking the electron flow.

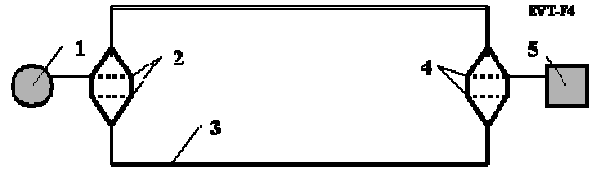

**Fig.3**. Example of electric line using the **AB-Tube** s. *Notations*: 1 – Electric constant-potential generator; 2 – acceleration nets; 3 – electric line from electronic tubes; 4 – brake nets; 5 – customer (user) of electric energy.

*Notes*: 1) We can change the charges in offered **AB-Tube:** Put into tube an ION gas and block the internal charge by an opposed outer charge. In this case we can transfer matter by the offered **AB-Tube.**

2) The electron gas has a very small density and produces significant pressure (the electrostatic force (pressure) is more by thousands times than gas pressure!). This electric charge may be active and effect force over a long distance.

That way the **AB-Tube** has a maximal lift force in the Earth atmosphere in comparison of the helium or hydrogen air balloons.

**Innovations**. The offered AB electron tube has principal differences from the used cathode-anode lamp (rectifier):

1) **AB-Tube** has permanent electronic gas sealed in the tube (high negative charge inside tube);
2) **AB-Tube** has positive electric charge in the outer surface of tube, which neutralizes the internal negative charge into tube.
3) **AB-Tube** does not have a hot cathode and anode, which are wasteful of power.
4) **AB-Tube** can have any length (up some thousands km). The cathode lamp has distance between the cathode and anode equal to some millimeters. They cannot work if this distance is great.
5) Cathode tubes continually inject the electrons from cathode and adsorb them by anode. The **AB-Tube** has permanent electron gas.
6) Electric **current** in cathode tube can flow only from cathode to anode and flow only when cathode is hot. Electric **current** into **AB-Tube** can flow in any direction and with any temperature of tube parts.
7) If we change the electronic gas inside the **AB-Tube** with an ion gas, we can transfer the matter by offered tube. This is not a possibility in a classical cathode tube.
8) The offered **AB-Tube** has very high conductivity in comparison with a cathode tube.

**Applications**.

There are numerous applications of offered **AB-Tube**s. Some of them are below:
1. Transfer electric energy in a long distance (up 10,000 km) with a small electric loss.
2. Superconductivity or close to superconductivity, 'Quasi-superconductivity'. The offered **AB-Tube** may have a very low electric resistance for any temperature because the electrons in the **AB-Tube** do not have large numbers of encounters with ions and do not lose energy from impacts with ions. The impact of electron to electron does not change the total impulse (momentum) of electron pairs and electron flow. If this idea is proved in experiments, that will be a big breakthrough in many fields of technology.

3. Cheap electric lines suspended at a high altitude (because the **AB-Tube**s have lift force in atmosphere and do not need electric masts and ground support paraphernalia]) (Fig.4a).
4. The big diameter **AB-Tube**s (including the electric lines) can be used as tramway for transportation (fig. 4a).
5. **AB-Tube**s can be used as vacuum tubes for an exit from the Earth's surface to outer space (a direct conduit outside of Earth's atmosphere from the surface!) (fig.4c). That may be used by Earth telescopes for observation of sky without atmospheric hindrances, or sending a plasma beam to space ships without atmospheric hindrances [8-9].
6. Transfer the electric energy from continent to continent through the Earth's ionosphere (fig. 4d) [11, 15].
7. Air balloons and dirigibles without expensive lifting gas.
8. Inserting an anti-gravitator cable into a vacuum-enclosing **AB-Tube** for near-complete elimination of air friction [4-5]. Same application for transmission of mechanical energy for long distance with minimum friction and losses. [4].
9. Increasing by some times the range of a conventional gun. They can shoot through the vacuum tube (to top of thinner layer of the atmosphere, up to 4-6 km) and projectile will fly in rare atmosphere where the air drag is small.

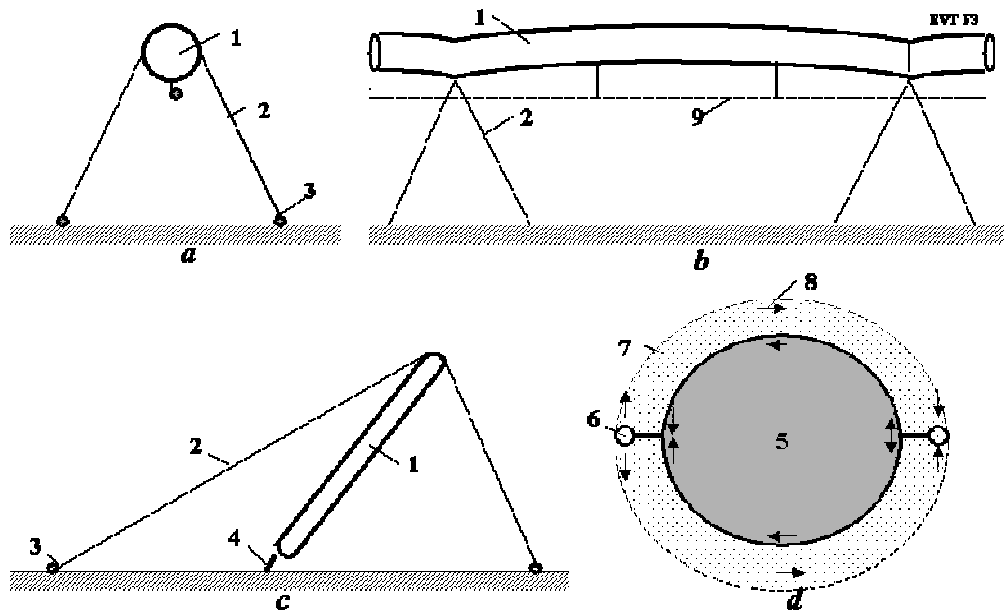

**Fig. 4**. Some application of **AB-Tube**s. *a-b*) High efficiency electric long distance lines or (also) tramway (*a* - front view, *b* – side view). *c*) Earth's telescope without atmosphere hindrances, or plasma beam without atmosphere hindrances, *d*) Transfer of long distance of electric energy through Earth ionosphere. *Notations*: 1 – electronic tube; 2 – guy lines; 3 – winch; 4 – telescope, or plasma beam injector, or gun; 5 – Earth; 6 – high altitude (100 km) electrostatic tower; 7 – ionosphere; 8 – electric current; 9 – tramway.

**Attachment**: In [3] author offered and applied this idea toward research the super high space tower (up to GEO, altitude 37,000 km). Below is a base for it and some results.

*1. Electron gas and space tower.* The electron gas consists of conventional electrons. In contrast to molecular gas, the electron gas has many surprising properties. For example, electron gas (having same mass density) can have different pressure in the given volume. Its pressure depends from electric intensity, but electric intensity is different in different part of given volume (fig.2a). For example, in our tube the electron intensity is zero in center of cylindrical tube and maximum near and at the tube surface.

The offered **AB-Tube** is the main innovation in the suggested tower. It has positive control charges, isolated thin film cover, and electron gas inside. The positive cylinder creates a net zero electric field inside the tube and electrons conduct themselves as conventional molecules that are equal in mass density at any point. When kinetic energy of electrons is less than the energy of negative ionization of the dielectric cover; or the material of the electric cover does not accept the negative ionization; the electrons are reflected from the cover. In the other case the internal cover layer is saturated by negative ions and begins also to reflect electrons. Important also is that the offered AB electrostatic tube has a neutral summary charge in outer space.

   *Advantages of electrostatic tower*. The offered electrostatic tower has very important advantages in comparison with conventional space elevator architecture commonly pictured today:
1. Electrostatic AB tower (mast) may be built from Earth's surface without rockets. That fact alone decreases the cost of a electrostatic mast by thousands of times.
2. One can have any height and payload capacity.
3. In particle, electrostatic tower can have the height of a geosynchronous orbit (37,000 km) WITHOUT the additional mandatory multi-hundred ton counterweight of a 'conventional' space elevator (and the long tether it must be suspended by to 120,000 - 160,000 km) [4], Ch.1.
4. The offered mast has less total mass by tens of times than conventional space elevator.
5. The offered mast can be built from less strong material than space elevator cable (see the computation here and in [4] Ch.1).
6. The offered tower can have high-speed electrostatic climbers impelled by high voltage electricity from Earth's surface [10, 6].
7. The offered tower is safer against the expected onslaught of meteorites (reentering space junk particles) than the cable of a 'conventional' space elevator, because any small meteorite damaging the cable may lead to a runaway collapse catastrophe for a space elevator, but it only creates small holes in a electrostatic tower. The electron escape may be compensated by automatic electron injection.
8. The electrostatic mast can bend in a needed direction when we give the appropriate electric voltage in needed parts of the mast.

   The electrostatic tower of heights up to 100 - 500 km may be built from current artificial fiber material available at the present time. The geosynchronous electrostatic tower needs stronger material having a strong coefficient $K \geq 2$ (whiskers or nanotubes, see below).

*2. Other applications of the* **AB-Tube**.

The offered **AB-Tube** with the positive charged cover and the electron gas inside may find the many applications in other technical fields. For example:
1) *Air dirigible*. (1) The airship from the thin film filled by an electron gas has 30% more lift force than conventional dirigible filled by costly helium. (2) An electron dirigible is significantly cheaper than the equivalent helium-filled dirigible because the helium is a very expensive gas (~$7/kg) and is becoming more so over time. (3) One does not have a problem with changing the lift force for fine control because it is no problem to add or to delete electrons.
2) *Long arm*. The offered electron control tube can be used as a long control work arm for performing work on a planet or other celestial body, rescue operations, repairing of other spaceships and so on. For details see [4] Ch.9.
3) *Super-reflectivity*. If free electrons are located between two thin transparent plates, that may be usable as a super-reflectivity mirror for a widely spectrum of radiation. That is necessary in many important technical fields as a light engine, multi-reflection propulsion [4] Ch.12 and thermonuclear power [16].

Other applications of electrostatic technology are Electrostatic solar wind propulsion [4] Ch.13, Electrostatic utilization of asteroids for space flight [4] Ch.14, Electrostatic levitation on the Earth and artificial gravity for spaceships and asteroids [4, 5 Ch.15], Electrostatic solar sail [3] Ch.18, Electrostatic space radiator [4] Ch.19, Electrostatic AB-Ramjet space propulsion [4], etc.

# Theory and Computation

Below the interested reader may find the evidence of main equations, estimations, and computations.

**1. Relation between the linear electric charge of tube and electron gas pressure on tube surface:**

$$p = \frac{\varepsilon_0 E^2}{2}, \quad E = k\frac{2\tau}{r}, \quad \varepsilon_0 = \frac{1}{4\pi k}, \quad \tau = \sqrt{\frac{2\pi r p}{k}}, \quad (1)$$

where $p$ is electron pressure, N/m$^2$; $\varepsilon_0 = 8.85 \times 10^{-12}$ F/m –electrostatic constant; $k = 9 \times 10^9$ Nm$^2$/C$^2$ is electrostatic constant; E is electric intensity, V/m; $\tau$ is linear charges of tube, C/m; $r$ is radius of tube, m.

Example, for atmospheric pressure $p = 10^5$ N/m$^2$ we receive $E = 1.5 \times 10^8$ V/m, N/C, the linear charge $\tau = 0.00833r$ C/m.

**2. Density of electron (ion) in 1 m$^3$ in tube.**

$$n = \frac{\tau}{\pi r^2 e} = \frac{1}{2\pi ek}\frac{E}{r} = 1.1 \cdot 10^8 \frac{E}{r},$$

$$M_e = m_e n, \quad M_i = \mu n_p n, \quad \mu = \frac{m_i}{m_p} \quad (2)$$

where $n$ is charge (electron or ion) density, 1/m$^3$; $e = 1.6 \times 10^{-19}$ C is charge of electron; $m_e = 9.11 \times 10^{-31}$ is mass of electron, kg; $m_p = 1.67 \times 10^{-27}$ is mass of proton, kg; $M_e$ is mass density of electron, kg/m$^3$; $M_i$ is mass density of ion, kg/m$^3$.

For electron pressure 1 atm the electron density (number particles in m$^3$) is $n = 1.65 \times 10^{16}/r$.

**3. Electric resistance of AD-tube.** We estimate the friction of electron about the tube wall by gas-kinetic theory

$$F_B = \eta_B SV, \quad \eta_B = \frac{1}{6}\rho V, \quad \rho = m_e n,$$

$$\overline{F} = \frac{F}{S} = \frac{1}{6}m_e n V^2, \quad V = \frac{j}{en} = \frac{i}{en\pi r^2}, \quad (3)$$

where $F_B$ is electron friction, N; $\eta_B$ is coefficient of friction; $S$ is friction area, m$^2$; $V$ is electron speed, m/s; $\rho$ is density of electron gas, kg/m$^3$; $\overline{F}$ is relative electron friction, N/m$^2$ ; $j$ is **current** density, A/m$^2$.

**4. Electric loss**. The electric loss (power) into tube is

$$P_T = \overline{F}_B SV, \quad S = 2\pi r L, \quad P_T = \frac{1}{3}\pi m_e n r L V^3,$$

$$P_T = \frac{m_e}{3e^3\pi^2}\frac{i^3 L}{n^2 r^5} = 7.5 \cdot 10^{24}\frac{i^3 L}{n^2 r^5} \quad [W], \quad (4)$$

where $P_T$ is electric loss, W; $L$ is tube length, m; $i$ is electric current, A.

**5. Relative electric loss** is

$$\overline{P}_T = \frac{P_T}{P}, \quad P = iU, \quad \overline{P}_T = \frac{m_e}{3\pi^2 e^3}\frac{i^2}{n^2 r^5}\frac{L}{U} = 7.5 \cdot 10^{24}\frac{i^2}{n^2 r^5}\frac{L}{U} = 7.4 \cdot 10^{25}\frac{j^2}{n^2 r}\frac{L}{U}, \quad (5)$$

Compare the relative loss the offered electric (tube) line and conventional electric long distance line. Assume the electric line have length $L = 2000$ km, electric voltage $U = 10^6$ V, electric **current** $i = 300$ A, atmospheric pressure into tube. For offered line having tube $r = 1$ m the relative loss equals $\overline{P}_T = 0.005$. For conventional electric line having cross section **copper** wire 1 cm$^2$ the relative loss is $\overline{P}_T = 0.105$. That is in 21 times more than the offered electric line. The computation of Equation (5) for atmospheric pressure and for ratio $L/U = 1$ are presented in fig. 5. As you see for electric line $L = 1000$ km, voltage $U = 1$ million V, tube radius $r = 2.2$ m, the electric **current** $i = 50$ A, the relative loss of electric power is one/millionth ($10^{-6}$), (only 50 W for transmitted power 50 millions watt!).

Moreover, the offered electric line is cheaper by many times, may be levitated into the atmosphere at high altitude, does not need a mast and ground, doesn't require expensive copper, does not allow easy surface access to line tapping thieves who wish to steal the electric energy. And this levitating electric line may be suspended with equal ease over sea as over land.

**6. Lift force of tube** ($L_{F,1}$, kg/m) and mass of 1 m length of tube ($W_1$. kg/m) is

$$L_{F,1} = \rho v = \rho \pi r^2, \quad W_1 = 2\pi r^2 \gamma \delta, \qquad (6)$$

where $\rho$ is air density, at sea level $\rho = 1.225$ kg/m$^3$; $v$ is volume of 1 m of tube length, m$^3$; $\gamma$ is density of tube envelope, for most plastic $\gamma = 1500 \div 1800$ kg/m$^3$; $\delta$ is film thickness, m.

*Example*. For $r = 10$ m and $\delta = 0.1$ mm, the lift force is 384 kg/m and cover mass is 11.3 kg/m.

**7. Artificial fiber and tube (cable) properties** [18]-[21]. Cheap artificial fibers are currently being manufactured, which have tensile strengths of 3-5 times more than steel and densities 4-5 times less than steel. There are also experimental fibers (whiskers) that have tensile strengths 30-100 times more than steel and densities 2 to 5 times less than steel. For example, in the book [18] p.158 (1989), there is a fiber (whisker) $C_D$, which has a tensile strength of $\sigma = 8000$ kg/mm$^2$ and density (specific gravity) of $\gamma = 3.5$ g/cm$^3$. If we use an estimated strength of 3500 kg/mm$^2$ ($\sigma = 7 \cdot 10^{10}$ N/m$^2$, $\gamma = 3500$ kg/m$^3$), than the ratio is $\gamma/\sigma = 0.1 \times 10^{-6}$ or $\sigma/\gamma = 10 \times 10^6$.

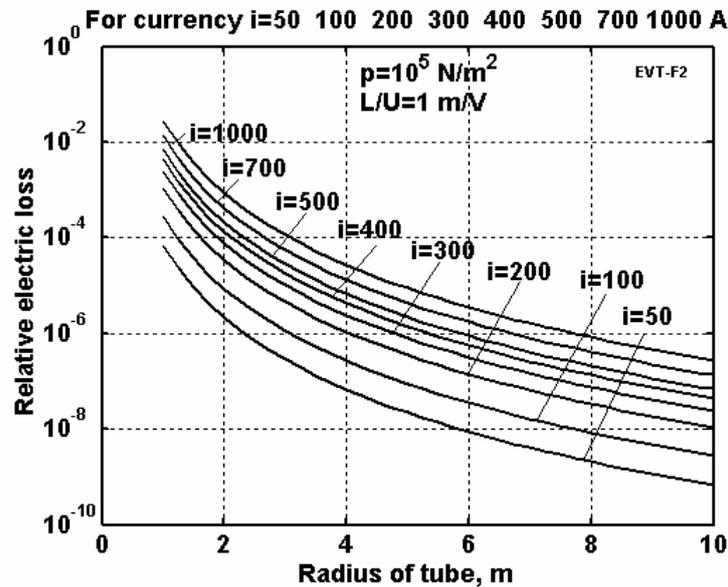

**Fig. 5**. Relative electric loss via radius of tube for electric **current** $i = 50 \div 1000$ A, the atmospheric pressure into tube and ratio $L/U = 1$.

Although the described (1989) graphite fibers are strong ($\sigma/\gamma = 10 \times 10^6$), they are at least still ten times weaker than theory predicts. A steel fiber has a tensile strength of 5000 MPA (500 kg/sq.mm), the theoretical limit is 22,000 MPA (2200 kg/mm$^2$) (1987); polyethylene fiber has a tensile strength 20,000 MPA with a theoretical limit of 35,000 MPA (1987). The very high tensile strength is due to its nanotube structure [21].

Apart from unique electronic properties, the mechanical behavior of nanotubes also has provided interest because nanotubes are seen as the ultimate carbon fiber, which can be used as reinforcements in advanced composite technology. Early theoretical work and recent experiments on individual nanotubes (mostly MWNT's, Multi Wall Nano Tubes) have confirmed that nanotubes are one of the stiffest materials ever made. Whereas carbon-carbon covalent bonds are one of the strongest in nature, a structure based on a perfect arrangement of these bonds oriented along the axis of nanotubes would produce an exceedingly strong material. Traditional carbon fibers show high strength and stiffness, but fall far short of the theoretical, in-plane strength of graphite layers by an order of magnitude. Nanotubes come close to being the best fiber that can be made from graphite.

For example, whiskers of Carbon nanotube (CNT) material have a tensile strength of 200 Giga-Pascals and a Young's modulus over 1 Tera Pascals (1999). The theory predicts 1 Tera Pascals and a Young's modules of 1-5 Tera Pascals. The hollow structure of nanotubes makes them very light (the specific density varies from 0.8 g/cc for SWNT's (Single Wall Nano Tubes) up to 1.8 g/cc for MWNT's, compared to 2.26 g/cc for graphite or 7.8 g/cc for steel). Tensile strength of MWNT's nanotubes reaches 150 GPa.

In 2000, a multi-walled carbon nanotube was tested to have a tensile strength of 63 GPa. Since carbon nanotubes have a low density for a solid of 1.3-1.4 g/cm³, its **specific strength** of up to 48,000 kN·m/kg is the best of known materials, compared to high-carbon steel's 154 kN·m/kg.

The theory predicts the tensile stress of different types of nanotubes as: Armchair SWNT - 120 GPa, Zigzag SWNT – 94 GPa.

Specific strength (strength/density) is important in the design of the systems presented in this paper; nanotubes have values at least 2 orders of magnitude greater than steel. Traditional carbon fibers have a specific strength 40 times that of steel. Since nanotubes are made of graphitic carbon, they have good resistance to chemical attack and have high thermal stability. Oxidation studies have shown that the onset of oxidation shifts by about $100^0$ C or higher in nanotubes compared to high modulus graphite fibers. In a vacuum, or reducing atmosphere, nanotube structures will be stable to any practical service temperature (in vacuum up 2800 °C. in air up 750°C).

In theory, metallic nanotubes can have an electric current density (along axis) more than 1,000 times greater than metals such as silver and copper. Nanotubes have excellent heat conductivity along axis up 6000 W/m·K. Copper, by contrast, has only 385 W/m·K.

About 60 tons/year of nanotubes are produced now (2007). Price is about $100 - 50,000/kg. Experts predict production of nanotubes on the order of 6000 tons/year and with a price of $1 – 100/kg to 2012.

Commercial artificial fibers are cheap and widely used in tires and countless other applications. The authors have found only older information about textile fiber for inflatable structures (Harris J.T., Advanced Material and Assembly Methods for Inflatable Structures, AIAA, Paper No. 73-448, 1973). This refers to DuPont textile Fiber **B** and Fiber **PRD-49** for tire cord. They are 6 times strong as steel (psi is 400,000 or 312 kg/mm²) with a specific gravity of only 1.5. Minimum available yarn size (denier) is 200, tensile module is $8.8 \times 10^6$ (**B**) and $20 \times 10^6$ (**PRD-49**), and ultimate elongation (percent) is 4 (**B**) and 1.9 (**PRD-49**). Some data are in Table 1.

**Table 1.** Material properties

| Material | Tensile strength kg/mm² | Density g/cm³ | Fibers | Tensile strength kg/mm² | Density g/cm³ |
|---|---|---|---|---|---|
| Whiskers | | | | | |
| AlB$_{12}$ | 2650 | 2.6 | QC-8805 | 620 | 1.95 |
| B | 2500 | 2.3 | TM9 | 600 | 1.79 |
| B$_4$C | 2800 | 2.5 | Allien 1 | 580 | 1.56 |
| TiB$_2$ | 3370 | 4.5 | Allien 2 | 300 | 0.97 |
| SiC | 1380-4140 | 3.22 | Kevlar or Twaron | 362 | 1.44 |
| **Material** | | | Dynecta or Spectra | 230-350 | 0.97 |
| Steel prestressing strands | 186 | 7.8 | Vectran | 283-334 | 0.97 |
| Steel Piano wire | 220-248 | | E-Glass | 347 | 2.57 |
| Steel A514 | 76 | 7.8 | S-Glass | 471 | 2.48 |
| Aluminum alloy | 45.5 | 2.7 | Basalt fiber | 484 | 2.7 |
| Titanium alloy | 90 | 4.51 | Carbon fiber | 565 | 1,75 |
| Polypropylene | 2-8 | 0.91 | Carbon nanotubes | 6200 | 1.34 |

Source: [15]-[18]. Howatsom A.N., Engineering Tables and Data, p.41.

Industrial fibers have up to $\sigma = 500 - 600$ kg/mm², $\gamma = 1500 - 1800$ kg/m³, and $\sigma/\gamma = 2,78 \times 10^6$. But we are projecting use in the present projects the cheapest films and cables applicable (safety $\sigma = 100 - 200$ kg/mm²).

**8. Dielectric strength of insulator.** As you see above, the tube needs film that separates the positive charges located in conductive layer from the electron gas located in the tube. This film must have a high dielectric strength. The current material can keep a high $E$ (see table 2 is taken from [10]).

Table 2. Properties of various good insulators (recalculated in metric system)

| Insulator | Resistivity Ohm-m. | Dielectric strength MV/m. $E_i$ | Dielectric constant, $\varepsilon$ |
|---|---|---|---|
| Lexan | $10^{17}$–$10^{19}$ | 320–640 | 3 |
| Kapton H | $10^{19}$–$10^{20}$ | 120–320 | 3 |
| Kel-F | $10^{17}$–$10^{19}$ | 80–240 | 2–3 |
| Mylar | $10^{15}$–$10^{16}$ | 160–640 | 3 |
| Parylene | $10^{17}$–$10^{20}$ | 240–400 | 2–3 |
| Polyethylene | $10^{18}$–$5\times10^{18}$ | 40–680* | 2 |
| Poly (tetra-fluoraethylene) | $10^{15}$–$5\times10^{19}$ | 40–280** | 2 |
| Air (1 atm, 1 mm gap) | | 4 | 1 |
| Vacuum ($1.3\times10^{-3}$ Pa, 1 mm gap) | | 80–120 | 1 |

*For room temperature 500 – 700 MV/m.
** 400–500 MV/m.

*Sources:* Encyclopedia of Science & Technology (New York, 2002, Vol. 6, p. 104, p. 229, p. 231) and Kikoin [17] p. 321.

*Note:* Dielectric constant $\varepsilon$ can reach 4.5 - 7.5 for mica ($E$ is up 200 MV/m), 6 -10 for glasses ($E$ = 40 MV/m), and 900 - 3000 for special ceramics (marks are CM-1, T-900) [17], p. 321, ($E$ =13 -28 MV/m). Ferroelectrics have $\varepsilon$ up to $10^4$ - $10^5$. Dielectric strength appreciably depends from surface roughness, thickness, purity, temperature and other conditions of materials. Very clean material without admixture (for example, quartz) can have electric strength up 1000 MV/m. As you see, we have the needed dielectric material, but it is necessary to find good (and strong) isolative materials and to research conditions which increase the dielectric strength.

**9. Tube cover thickness**. The thickness of the tube's cover may be found from Equation

$$\delta = \frac{rp}{\sigma}, \qquad (7)$$

where $p$ is electron pressure minus atmospheric pressure, N/m$^2$. If electron pressure is little more then the atmospheric pressure the tube cover thickness may be very thin.

**10. Mass of tube cover**. The mass of tube cover is

$$M_1 = \delta\gamma, \quad M = 2\pi rL\gamma\delta, \qquad (8)$$

where $M_1$ is 1 m$^2$ cover mass, kg/m$^2$; $M$ is cover mass, kg.

**11. The volume $V$ and surface of tube $s$** are

$$V = \pi r^2 L, \quad s = \pi rL, \qquad (9)$$

where $V$ is tube volume, m$^3$; $s$ is tube surface, m$^2$.

**12. Relation between tube volume charge and tube liner charge** for neutral tube is

$$E_V = \frac{\rho r}{2\varepsilon_0}, \quad E_s = \frac{\tau}{2\pi\varepsilon_0 r}, \quad E_V = E_s, \quad \tau = \pi\rho r^2, \quad \rho = \frac{\tau}{\pi r^2}, \qquad (10)$$

where $\rho$ is tube volume charge, C/m$^3$; $\tau$ is tube linear charge, C/m.

**13. General charge of tube**. We got equation from

$$\tau = 2\pi\varepsilon\varepsilon_0 Er, \quad Q = \tau L, \quad Q = 2\pi\varepsilon\varepsilon_0 ErL, \qquad (11)$$

where $Q$ is total tube charge, C; $\varepsilon$ is dielectric constant (see Table 2).

**14. Charging energy**. The charged energy is computed by equation

$$W = 0.5QU, \quad U = \delta E, \quad W = 0.5Q\delta E, \qquad (12)$$

where $W$ is charge energy, J; $U$ is voltage, V.

15. **Mass of electron gas**. The mass of electron gas is

$$M_e = m_e N = m_e \frac{Q}{e}, \qquad (13)$$

where $M_e$ is mass of electron gas, kg; $m_e = 9.11 \times 10^{-31}$ kg is mass of electron; $N$ is number of electrons, $e = 1.6 \times 10^{-19}$ is the electron charge, C.

16. **Transfer of matter** (Matter flow of ion gas). If we change the electron gas by the ion gas, our tube transfer charged matter with very high speed

$$M = M_i \pi r^2 V, \quad M_i = \mu m_p n,$$
$$V = \frac{i}{en\pi r^2}, \quad M = \frac{m_p}{e}\mu = 1.04 \cdot 10^{-8} \mu i \qquad (14)$$

where $M$ is the mass flow, kg/s; $M_i$ is the gas ion density, kg/m³; $\mu = m_i/m_p$; $V$ is ions speed, m/s.

*Example*: We want to transfer to a remote location the nuclear breeder fuel – Uranium-238. ($\mu = 238$) by line having $i = 1000$ A, $r = 1$ m, ion gas pressure 1 atm. One day contains 86400 seconds.

The equation (12) gives $M = 214$ kg/day, speed $V = 120$ km/s. The AB-tubes are suitable for transferring small amounts of a given matter. For transferring a large mass the diameter of tube and electric current must be larger.

We must also have efficient devices for ionization and utilization of the de-ionization (recombination) energy.

The offered method allows direct conversion of the ionization energy of the electron gas or ion gas to light (for example, by connection between the electron and ion gases).

17. **Electron gas pressure**. The electron gas pressure may be computed by equation (1). This computation is presented in fig. 6.

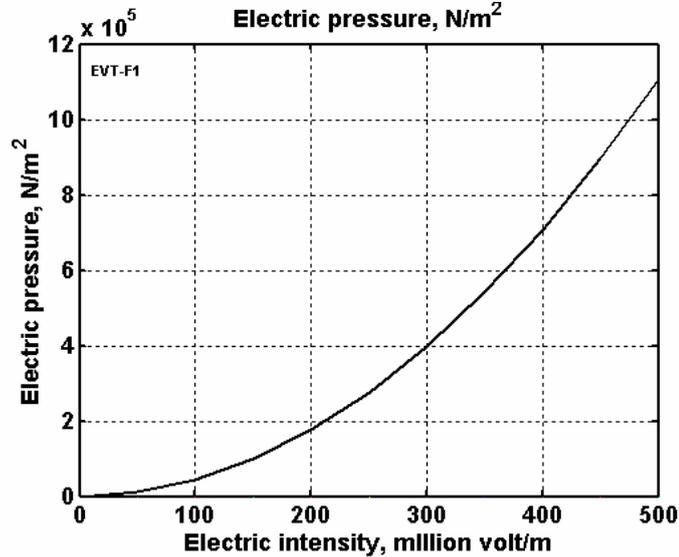

**Fig. 6.** Electron pressure versus electric intensity

As you see the electron pressure reaches 1 atm for an electric intensity 150 MV/m and for negligibly small mass of the electron gas.

18. **Power for support of charge**. Leakage current (power) through the cover may be estimated by equation

$$I = \frac{U}{R}, \quad U = \delta E = \frac{r\varepsilon_0 E}{\sigma}, \quad R = \rho\frac{\delta}{s}, \quad I = \frac{sE}{\rho}, \quad W_l = IU = \frac{\delta s E^2}{\rho}, \qquad (15)$$

where $I$ is electric **current**, A; $U$ is voltage, V; $R$ is electric resistance, Ohm; $\rho$ is specific resistance, Ohm·m; $s$ is tube surface area, m².

The estimation gives the support power has small value.

## Quasi-superconductivity of AB-Tube.

The proposed **AB-Tube may** become what we may term 'quasi-superconductive' when magnetic pressure equals electrostatic pressure. In this case electrons cannot contact with the tube wall, do not experience resistance friction and the **AB-Tube** thus experiences this 'quasi-superconductivity'. Let us to get this condition:

$$P_e = \frac{\varepsilon_0 E^2}{2}, \quad P_m = \frac{B^2}{2\mu_0}, \quad P_e = P_m, \quad c = \frac{1}{\sqrt{\mu_0 \varepsilon_0}}, \quad E = cB \ , \tag{16}$$

where $P_e$ is electronic pressure, N/m$^2$; $P_m$ is magnetic pressure, N/m$^2$; $B$ is magnetic intensity, T; $E$ is electric intensity, V/m; $c$ is light speed, $c = 3 \times 10^8$ m/s; $\varepsilon_0, \mu_0 = 4\pi \times 10^{-7}$ are electrostatic and magnetic constants. The relation $E = cB$ is important result and condition of tube superconductivity. For electron pressure into tube 1 atm, the $E = 1.5 \times 10^8$ V/m (see above) and $B = 0.5$ T.

From Eq. (16) we receive the relation between the electric **current** and the tube charge for **AB-Tube** 'quasi-superconductivity' as

$$E = cB, \quad E = \frac{1}{2\pi\varepsilon_0}\frac{\tau}{r}, \quad B = \frac{\mu_0 i}{2\pi r}, \quad i = c\tau \ , \tag{17}$$

where $i$ is electric **current**, A; $\tau$ is liner charge of tube, C/m.

For electron pressure equals 1 atm and $r = 1$m the linear tube charge is $\tau = 0.00833$ C/m (see above) and the request electric **current** is $i = 2.5 \times 10^6$ A ($j = 0.8$ A/m$^2$). For $r = 0.1$ m the **current** equals $i = 2.5 \times 10^5$ A. And for r = 0.01 m the **current** equals $i = 2.5 \times 10^4$ A.

Unfortunately, the requested electron speed (for true and full normal temperature 'superconductivity') equals light speed $c$. That means we cannot exactly reach it, but we can came very close and we can have very low electric resistance of **AB-Tube.**

**Information about high speed of electron and ion beam.** Here $\gamma = (1-\beta^2)^{-1/2}$ is the relativistic scaling factor; quantities in analytic formulas are expressed in SI or cgs units, as indicated; in numerical formulas $I$ is in amperes (A), **B** is in gauss (G, 1 T = 10$^4$ G), electron linear density $N$ is in cm$^{-1}$, temperature, voltage, and energy are in MeV, $\beta_z = v_z/c$, and $k$ is Boltzmann's constant.

For computation electrostatic and magnetic fields about light speed are useful the equations of relativistic theory (Lorenz's Equations):

$$\begin{aligned} E_x &= E_{1x}, & H_x &= H_{1x}, \\ \sqrt{1-\beta^2}\,E_y &= E_{1y} - vB_{1z}, & \sqrt{1-\beta^2}\,H_y &= H_{1y} + vD_{1z}, \\ \sqrt{1-\beta^2}\,E_z &= E_{1z} + vB_{1y}, & \sqrt{1-\beta^2}\,H_z &= H_{1z} - vD_{1y}, \end{aligned} \tag{18}$$

where lower index "$_1$" means the immobile system coordinate, $E$ is electric intensity, V/m; $H$ is magnetic intensity, A/m; $v$ is speed of mobile system coordinate along axis $x$, m/s; $D$ is electric displacement. C/m$^2$; $\beta = v/c$ is relative speed one system about the other.

Relativistic electron gyroradius [22]:

$$r_e = \frac{mc^2}{eB}(\gamma^2 - 1)^{1/2} \ \ (cgs) = 1.70 \cdot 10^3 (\gamma^2 - 1)^{1/2} B^{-1} \ \ \text{cm}. \tag{19}$$

Relativistic electron energy:

$$W = mc^2\gamma = 0.511\gamma \quad \text{MeV}. \tag{20}$$

Bennett pinch condition:

$$I^2 = 2Nk(T_e + T_i)c^2 \ \ (cgs) = 3.20 \cdot 10^{-4} N(T_e + T_i) \quad \text{A}^2. \tag{21}$$

Alfven-Lawson limit:

$$I_A = (mc^3/e)\beta_z\gamma \ \ (cgs) = (4\pi mc/\mu_0 e)\beta_z\gamma \ \ (SI) = 1.70 \cdot 10^4 \beta_z\gamma \quad \text{A}. \tag{22}$$

The ratio of net current to $I_A$ is

$$\frac{I}{I_A} = \frac{\nu}{\gamma}. \tag{23}$$

Here $\nu = N r_e$ is the Budker number, where $r_e = e^2/mc^2 = 2.82 \cdot 10^{-13}$ cm is the classical electron radius. Beam electron number density is

$$n_b = 2.08 \cdot 10^8 J \beta^{-1} \quad \text{cm}^{-3}, \tag{24}$$

where J is the current density in A cm-2. For a uniform beam of radius a (in cm):

$$n_b = 6.63 \cdot 10^7 I a^{-2} \beta^{-1} \quad \text{cm}^{-3} \tag{25}$$

and

$$\frac{2 r_e}{a} = \frac{\nu}{\gamma}, \tag{26}$$

Child's law: nonrelativistic space-charge-limited current density between parallel plates with voltage drop $V$ (in MV) and separation $d$ (in cm) is

$$J = 2.34 \cdot 10^3 V^{3/2} d^{-2} \quad \text{A cm}^{-2} \tag{27}$$

The condition for a longitudinal magnetic field $B_z$ to suppress filamentation in a beam of current density $J$ (in A cm$^{-2}$) is

$$B_z > 47 \beta_z (\gamma J)^{1/2} \quad \text{G}. \tag{28}$$

Kinetic energy necessary to accelerate a particle is

$$K = (\gamma - 1) mc^2. \tag{29}$$

The de Broglie wavelength of particle is $\lambda = h/p$, where $h = 6.6262 \times 10^{-34}$ J·s is Planck constant, $p$ is particle momentum. Classical radius of electron is $2.8179 \times 10^{-15}$ m.

## Conclusion

The offered inflatable electrostatic AB tube has indisputably remarkable operational advantages in comparison with the conventional electric lines.

The main innovations and applications of **AB-Tube**s are:
1. Transferring electric energy in a long distance (up 10,000 km) with a small electric loss.
2. 'Quasi-superconductivity'. The offered **AB-Tube** may have a very low electric resistance for any temperature because the electrons in the tube do not have ions and do not lose energy by impacts with ions. The impact the electron to electron does not change the total impulse (momentum) of couple electrons and electron flow. If this idea is proved in experiment, that will be big breakthrough in many fields of technology.
3. Cheap electric lines suspended in high altitude (because the **AB-Tube** can have lift force in atmosphere and do not need ground mounted electric masts and other support structures) (Fig.4a).
4. The big diameter **AB-Tube**s (including the electric lines for internal power can be used as tramway for transportation (fig. 4a).
5. **AB-Tube** s can be used as vacuum tubes for an exit from the Earth's surface to outer space (out from Earth's atmosphere) (fig.4c). That may be used by an Earth telescope for observation of sky without atmosphere hindrances, or sending of a plasma beam to space ships without atmosphere hindrances [8-9].
6. Transfer of electric energy from continent to continent through the Earth's ionosphere (fig. 4d) [11, 15].
7. Inserting an anti-gravitator cable into a vacuum-enclosing **AB-Tube** for near-complete elimination of air friction [4-5]. Same application for transmission of mechanical energy for long distances with minimum friction and losses. [4].
8. Increasing in some times the range of a conventional gun. They can shoot through the vacuum tube (up 4-6 km) and projectile will fly in the rare atmosphere where air drag is small.
9. Transfer of matter a long distance with high speed (including in outer space, see other of author's works).
10. Interesting uses in nuclear and high energy physics engineering (inventions).

The offered electronic gas may be used as filling gas for air balloons, dirigibles, energy storage, submarines, electricity-charge devices (see also [1]-[17, 23]).

## Acknowledgement

The author wishes to acknowledge Joseph Friedlander for helping to correct the author's English and for useful technical suggestions.